\documentclass[aps,prb,twocolumn,nopacs,amsmath,superscriptaddress,floatfix]{revtex4}

\usepackage{amsmath}
\usepackage{amsfonts}
\usepackage{epsfig}
\usepackage{graphicx}
\usepackage{graphics}
\usepackage{url}
\usepackage[english]{babel}
\usepackage{dcolumn}
\usepackage{color}
\begin{document}

\title{Fluctuation properties of acoustic phonons generated by ultrafast optical excitation of a quantum dot}

\author{D.~Wigger}
\affiliation{Institut f\"ur Festk\"orpertheorie, Universit\"at M\"unster, Wilhelm-Klemm-Str.~10, 48149
M\"unster,
Germany}

\author{D.~E.~Reiter}
\affiliation{Institut f\"ur Festk\"orpertheorie, Universit\"at M\"unster, Wilhelm-Klemm-Str.~10, 48149
M\"unster,
Germany}

\author{V.~M.~Axt}
\affiliation{Theoretische Physik III, Universit\"at Bayreuth, 95440
Bayreuth,
Germany}

\author{T.~Kuhn}
\affiliation{Institut f\"ur Festk\"orpertheorie, Universit\"at M\"unster, Wilhelm-Klemm-Str.~10, 48149
M\"unster,
Germany}

\date{\today}

\begin{abstract}
We study theoretically the fluctuation properties of acoustic phonons created
in a semiconductor quantum dot after ultrafast optical excitation. An
excitation with a single ultrafast pulse creates an exciton confined to the
quantum dot, which is coupled to longitudinal acoustic phonons. This leads to
the formation of a polaron in the quantum dot accompanied by the emission of
a phonon wave packet. We show that the fluctuations of the lattice
displacement associated with the wave packet after a single laser pulse
excitation in resonance with the exciton transition are always larger than their 
respective vacuum values. Manipulating the exciton with a second pulse can result 
in a reduction of the fluctuations below their vacuum limit, which means that the phonons are
squeezed. We show that the squeezing properties of the wave packet strongly
depend on the relative phase and the time delay between the two laser pulses.
\end{abstract}


\maketitle

\section{Introduction}

The creation and manipulation of nonclassical quantum states of bosonic
systems continue to attract large interest. Prominent examples of such
nonclassical states are squeezed states. In a squeezed state the fluctuations
of a given variable fall below their corresponding vacuum value at the cost
of increased fluctuations of its conjugate variable in order to satisfy
Heisenberg's uncertainty relation. A well established field since many years
is squeezed light, which can be generated in nonlinear optical processes like
parametric down conversion \cite{wu1986gen} and which has applications in
optical communication and measurements.\cite{yamamoto1986pre,caves1981qua}

Phonons are another type of bosons where nonclassical states have become of
growing interest in the past years. Squeezed phonons have been the subject of
many experimental \cite{misochko2000pha, misochko2000pec,
johnson2009directly, garrett1997vac, hussain2010abs} and theoretical
\cite{hu1996squ, hu1997pho, sauer2010, reiter2011wig, reiter2011coh,
papenkort2012optical, daniels2011quantum} studies. In most cases the
considered phonons had a fixed frequency, either because they were optical
phonons \cite{misochko2000pha, misochko2000pec, johnson2009directly,
hu1996squ, hu1997pho, sauer2010, reiter2011wig, reiter2011coh,
papenkort2012optical} or a van Hove singularity appeared in the spectrum of
acoustic phonons.\cite{garrett1997vac} An indication for squeezing is then
the appearance of an oscillation with the double phonon frequency, however
this alone is not yet an unambiguous proof for squeezing.\cite{hussain2010abs, sauer2010}

Phonons with a fixed frequency do not travel in the crystal because they have
a vanishing group velocity. Therefore, squeezing produced in such kind of
phonon systems remains in the region where it has been generated. In
contrast, in the case of photons squeezing can be transported at the speed of
light from the place where it is generated to other places. Such a transport
of squeezing could also occur for squeezed states of acoustic phonons which,
due to their approximately linear dispersion relation travel through the
crystal at the speed of sound. An example for squeezing of acoustic phonons,
which has been studied in the past, is the phonon analog of parametric down
conversion, i.e., the decay of a coherent optical phonon into a pair of
acoustic phonons.\cite{hu1996squ,daniels2011quantum} Interestingly, however,
it turned out, while indeed squeezing of the lattice displacement can be
achieved in this case, the squeezing again remains where it has been
generated and does not travel even though the generated phonons lead to a 
finite transport of energy away from the dot. The reason is that here the squeezing is
strongly related to the quantum correlations between phonons with opposite
wave vectors generated in the decay process. In this paper we will study a
different process, the generation of acoustic phonons associated with the
ultrafast optical excitation and manipulation of an exciton in a quantum dot
(QD) structure.

Quantum dots with their discrete level structure of electronic excitations
are often referred to as artificial atoms. Due to their embedding in a
surrounding semiconductor matrix, however, they are much more strongly
coupled to the environment than real atoms. Most important for the dynamics
of the QD exciton is in many cases the coupling to longitudinal acoustic (LA)
phonons via the deformation potential interaction, leading to various, often
undesired phenomena like a phononic background in absorption or luminescence
spectra \cite{besombes2001acoustic,krummheuer2002theory} or a damping of Rabi
oscillations.\cite{ramsay2010damping,glassl2011long,krugel2005role} However,
this coupling also gives rise to interesting spatiotemporal dynamics of the
generated phonons, which typically consist of a localized part remaining in
the region of the QD and a traveling part leaving the QD. This latter part
may be reflected at a surface and reenter the QD \cite{krummheuer2005coupled}
or may travel to another QD influencing its optical properties.\cite{huneke2008impact} 
Also bolometric measurements can be used to monitor the spatio-temporal dynamics of
phonon wave packets.\cite{hawker1999energy, bellingham2001acoustic}

In previous studies we have analyzed the possibility of generating phonon
squeezing in the case of a QD coupled to longitudinal optical (LO) phonons.
\cite{sauer2010, reiter2011wig} We have shown that a single ultrafast excitation 
resonant to the exciton transition cannot create squeezing, while a resonant two pulse excitation 
can lead to squeezed phonons. For LO phonons with a single 
frequency it is possible to transform the coupling to a coupling with 
a single effective phonon mode. The
dynamics can then be well illustrated by means of the Wigner function.
\cite{reiter2011wig} Such a reduction is not possible in the case of
acoustic phonons with their continuum of phonon frequencies. It is the aim of
this paper to analyze the fluctuation properties of LA phonons generated
after ultrafast resonant excitation of the lowest QD exciton. In particular, we will show that indeed
squeezed LA phonons may be created and that squeezed phonons can be emitted
from the QD in the form of a phonon wave packet.

The paper is organized as follows. In Sec.~\ref{s:model} we introduce the
theoretical model of the QD interacting with light pulses and LA phonons and
we define the relevant variables for the study of phonon squeezing.
Section~\ref{s:results} is then devoted to the results of our calculations,
where we first consider the case of excitation by a single pulse, which turns
out not to produce squeezing, and then turn to a two-pulse excitation, where
under suitable conditions squeezing is found. Finally, in
Sec.~\ref{s:conclus} we finish with some concluding remarks.

\section{Model system}\label{s:model}

We consider a QD in the strong confinement limit. Assuming excitations by
circularly polarized light, we can restrict the electronic states to a
two-level system consisting of the ground state $|g\rangle$ and the lowest
exciton state $|x\rangle$. These states are coupled to a classical light
field as well as to LA phonons. The Hamiltonian of the system then reads
\begin{eqnarray}
\hat{H}  & = &
\hbar\big[\Omega
+ \sum_{{\bf q}}
(g_{{\bf q}} \,\hat b_{{\bf q}} + g_{{\bf q}}^*\,\hat b_{{\bf q}}^\dag )
\big] | x \rangle \langle x |
\nonumber \\ & &
+ \hbar \sum_{{\bf q}} \,\omega_{\bf q} \hat b_{{\bf q}}^{\dag}\hat b_{{\bf q}}
  - {\bf \hat{P}} \cdot {\bf E}\,,
\label{eq:hamilton}
\end{eqnarray}
where $\hbar \Omega$ is the exciton energy, $\hat b_{{\bf q}}^\dag$ ($\hat
b_{{\bf q}}$) denotes the creation (annihilation) operator of a LA phonon
with wave vector ${\bf q}$, $\omega_{\bf q}=cq$ is the phonon dispersion
relation with the longitudinal sound velocity $c$, and $g_q$ is the
electron-phonon coupling matrix element. We restrict ourselves to the case of
deformation potential interaction, which for typical InAs/GaAs quantum dots
has been found to be the dominant interaction mechanism on a picosecond time
scale.\cite{vagov2004non} Assuming for simplicity a spherical QD geometry and a 
harmonic oscillator confinement, the coupling matrix element reads
\begin{equation}
g_q = \sqrt{\frac{1}{2\varrho \hbar V \omega_q}}q\left( D^e \mathrm e^{-\frac14q^2a_e^2}
- D^h \mathrm e^{-\frac14q^2a_h^2} \right)\,,
\label{eq:gq}
\end{equation}
with $\varrho$ being the crystal density and $V$ the normalization volume of
the crystal. $D^e$ ($D^h$) are the deformation potentials of electrons
(holes) and $a_e$ ($a_h$) the spatial widths of the electron (hole) wave
functions. The classical laser field ${\bf E}$ is coupled to the polarization
$\hat{\bf P}={\bf M}_0|x\rangle \langle g|+{\bf M}^\ast_0|g\rangle \langle
x|$ with the dipole matrix element ${\bf M}_0$. We consider ultrafast laser pulses 
that are, however, spectrally narrow enough to realize a selective resonant excitation 
of the exciton. Once the selectivity has been accounted for by keeping only the 
resonantly coupled electronic levels the pulse duration is the shortest 
time scale in the problem and we can safely model $E(t)$ as a series of delta-functions. 
We take GaAs material parameters and a QD with
$L=a_e2\sqrt{\ln{2}}=5\,{\rm nm}$ diameter.\cite{krummheuer2005pure}

A basic quantity for the lattice dynamics is the lattice displacement
associated with LA phonons
\begin{equation}
\hat{ {\bf u} }({\bf r}) = - i \sum_{\bf q} \sqrt{\frac{\hbar}{2\varrho V \omega_{\bf q}}}
(\hat{ b}_{\bf q} \mathrm e^{i{\bf q}\cdot{\bf r}} -
\hat{ b}_{\bf q}^\dag \mathrm e^{-i{\bf q}\cdot{\bf r}})
\frac{{\bf q}}{q} . \label{eq:u}
\end{equation}
In this paper we are particularly interested in the fluctuation squares of the
lattice displacement, $(\Delta u({\bf r}))^2=\langle \hat{u}({\bf r})^2
\rangle - \langle \hat{u}({\bf r})\rangle^2$. Assuming a spherical QD, all
quantities depend only on the distance $r$ from the QD center. For the
interpretation of the results we introduce the relative excitation induced
fluctuations squares
\begin{equation}
D_u(r,t) = \frac{(\Delta u(r,t))^2 - (\Delta u_{\rm vac} )^2}{(\Delta u_{\rm vac})^2}\,,
\end{equation}
where $(\Delta u_{\rm vac})^2$ are the fluctuation squares of the lattice
displacement in the phonon vacuum state. For simplicity we will refer to $D_u$ 
just as fluctuations in the following. We call a state ``squeezed'' if
$(\Delta u)^2<(\Delta u_{\rm vac})^2$. Thus squeezing manifests itself
directly by fluctuations $D_u(r)<0$. For LA phonons the vacuum
fluctuation $(\Delta u_{\rm vac})^2$ is calculated by taking the definition
according to Eq.~\eqref{eq:u} and integrating over the whole Brillouin 
zone assumed as spherical as in the Debey model,
i.e., up to the maximal wave vector $q_{\rm max}=(24\pi^2)^{1/3}/a$ for a zincblende 
structure with the lattice constant $a$. One finds
\begin{equation}
(\Delta u_{\rm vac})^2 = \frac{\hbar}{2\varrho c a^2}\left(\frac{3}{\pi}\right)^{2/3}
\approx 5.83\cdot 10^{-6}\,\rm{nm}^2 .
\end{equation}

For completeness we mention that the vacuum uncertainty of the lattice
momentum $(\Delta\pi_{\rm vac})^2$ for LA phonons is given by
\begin{equation}
(\Delta \pi_{\rm vac})^2 = \frac{\hbar\varrho c\pi^2a^2}{16}\left(\frac{3}{\pi}\right)^{4/3}
\approx 2.09\cdot 10^{4}\,\left(\frac{\rm meV\,ps}{\rm{nm}}\right)^2
\end{equation}
showing that for LA phonons the uncertainty product is
\begin{equation}
(\Delta u_{\rm vac})^2(\Delta \pi_{\rm vac})^2 = \frac{9}{8} \frac{\hbar^2}{4}
\end{equation}
and thus exceeds the Heisenberg limit already in the ground state.

To calculate the expectation values of the lattice displacement and its
fluctuations, we need expectation values of the types $\langle b_q\rangle$,
$\langle b_qb_{q^\prime}\rangle$ and $\langle b^\dag_qb_{q^\prime}\rangle$.
For the case of excitation by ultrafast laser pulses analytical results
for these quantities can be found within a generating function formalism,\cite{axt2005pho} 
which will form the basis for the analytical expressions given in Sec. \ref{s:results}.

\section{Results}\label{s:results}

\subsection{Single pulse excitation}

\begin{figure}[ht]
\includegraphics[width=\columnwidth]{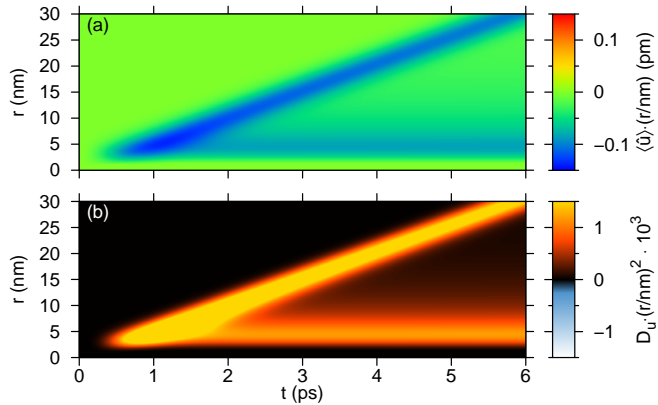}
\caption{\label{fig:1puls} (Color online) (a) Expectation value $\langle
\hat{u} \rangle\cdot r$ and (b) fluctuations $D_u\cdot r^2$ as functions of time $t$
and distance $r$ from the QD center after the excitation with a single
ultrafast laser pulse of pulse area $\pi/2$ at $t=0$. }\end{figure}

Let us start by looking at the phonon dynamics after excitation by a single
ultrafast laser pulse. In our previous study of the fluctuation dynamics of
LO phonons \cite{reiter2011wig} we have seen that the most interesting
phenomena appear in the case of excitation by pulses with a pulse area of
$\pi/2$. It turns out that the same holds for LA phonons. Therefore, in this
paper we will restrict our analysis to such pulses. We want to mention, that
analytical results for the phonon dynamics like the ones shown here can be
derived for any pulse area.

An excitation with a single laser pulse of pulse area $\pi/2$ creates an
equal superposition of ground state $|g\rangle$ and exciton $|x\rangle$ in
the electronic part of the system. The creation of the exciton causes a
change in the charge carrier distributions in the QD. Due to the deformation
potential interaction this gives rise to a local deformation of the lattice
corresponding to a shift of the equilibrium positions of the lattice ions. In
the case of acoustic phonons, this instantaneous shift in the equilibrium
positions leads to the creation of a localized polaron in the QD and a wave
packet leaving the QD.\cite{krummheuer2005pure} The expectation value of the
lattice displacement after the pulse reads
\begin{equation}
\langle \hat{u}(r,t) \rangle = u_0 U(r,t) =u_0 [ W(r,t) + P(r) ] \label{eq:u1}
\end{equation}
with $u_0=\frac{1}{8\varrho c^2\pi^2}$. $U(r,t)$ denotes the normalized
lattice displacement created by a single pulse excitation which can be
decomposed into the part $W(r,t)$ describing the wave packet leaving the QD
given by
\begin{eqnarray}
W(r,t) &=& - \frac{1}{r}\Big[ D^e\frac{\sqrt{\pi}}{a_e}\mathrm e^{-\frac{(r-ct)^2}{a_e^2}}
- D^h\frac{\sqrt{\pi}}{a_h}\mathrm e^{-\frac{(r-ct)^2}{a_h^2}} \Big] \label{eq:W}\\
&+& \frac{2}{\pi r^2} \Big[ D^e\operatorname{erf}\left(\frac{r-ct}{a_e}\right)
- D^h\operatorname{erf}\left(\frac{r-ct}{a_h}\right) \Big] \notag \\
&-& \frac{1}{r}\Big[ D^e\frac{\sqrt{\pi}}{a_e}\mathrm e^{-\frac{(r+ct)^2}{a_e^2}}
- D^h\frac{\sqrt{\pi}}{a_h}\mathrm e^{-\frac{(r+ct)^2}{a_h^2}} \Big] \notag\\
&+& \frac{2}{\pi r^2} \Big[ D^e\operatorname{erf}\left(\frac{r+ct}{a_e}\right)
- D^h\operatorname{erf}\left(\frac{r+ct}{a_h}\right) \Big] \notag
\end{eqnarray}
and $P(r)=- W(r,t=0)$ describing the localized polaron in the QD. Here,
$\operatorname{erf}(x)$ denotes the error function. For the wave packet the
dominant part for $t\gg L/c$ is the first line in Eq.~(\ref{eq:W}) that is
a Gaussian centered around $r=ct$ and decaying $\sim 1/r$. The other terms
are decaying with $1/r^2$ or describe an incoming wave packet that only
contributes for small $t$. For small $t$, however, and thus also for the
calculation of the polaron part $P(r)$ all terms of Eq.~(\ref{eq:W}) must be
taken into account, because the divergences in front of the Gaussians and the
error functions compensate each other resulting in a finite value at $r=0$.

Figure \ref{fig:1puls}(a) shows the expectation value of the lattice
displacement $\langle \hat{u}\rangle$ multiplied by $r$ plotted as a function
of time $t$ after the laser pulse and distance $r$ from the QD center. In
this Figure the polaron can be identified as the horizontal line at $r\approx 4\,{\rm
nm}$. The phonon wave packet that leaves the QD with the sound velocity
$c\approx 5\,{\rm nm/ps}$ is seen as diagonal line. The amplitude of both
polaron and wave packet are negative.

In Fig.~\ref{fig:1puls}(b) the corresponding fluctuations $D_u$ multiplied by $r^2$
are plotted as a function of time $t$ and position $r$. It is clearly seen
that $D_u$ is restricted to the same space-time regions where also the
displacement is non-vanishing. Thus, the two lines in the plot can be
identified as fluctuations of the polaron and of the traveling wave packet. Respectively
$D_u$ has the same shape as the displacement. This is confirmed by the
calculations showing that for a single pulse excitation there is always $D_u
\sim \langle \hat{u}\rangle^2$. Because $D_u$ is positive all the time we
find that, as in the case of LO phonons, a single ultrafast, resonant excitation of
the QD never creates squeezed phonons in this system.

\subsection{Two pulse excitation}

\begin{figure}[ht]
\includegraphics[width=0.7\columnwidth]{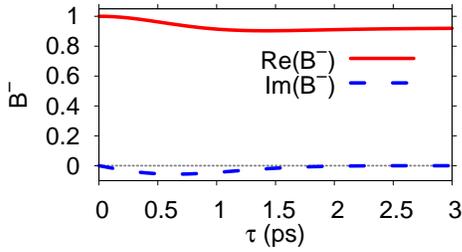}
\caption{(Color online) Real and imaginary part of the interaction amplitude
$B^-$ as a function of the delay time $\tau$.}
\label{fig:B}
\end{figure}

In the next step we consider an excitation with two laser pulses each with a
pulse area of $\pi/2$. The first pulse arrives at time $t=-\tau$ (with $\tau
\ge 0$). The second pulse arrives at $t=0$ and has a relative phase of $\phi$
with respect to the first pulse. The expectation value of the displacement
for $t>0$ after this two-pulse excitation can be written as
\begin{equation}
\langle \hat{u}(r,t)\rangle = u_0 [ U(r,t+\tau) + 
\operatorname{Re}\left(B^-(\tau)\mathrm e^{i\phi}\right) U(r,t) ] , \label{eq:u2}
\end{equation}
where $U(r,t)$ is the normalized lattice displacement after a single pulse
[see Eqs.~(\ref{eq:u1}), (\ref{eq:W})]. The first pulse exciting the system at
$t=-\tau$ gives rise to the displacement $U(r,t+\tau)$, the second pulse
arriving at $t=0$ then creates $U(r,t)$. The two terms are connected via the
interference amplitude
\begin{equation}
B^-(\tau) = \exp \left[ \sum_q \left|\frac{g_q}{\omega_q}\right|^2
\big( \mathrm e^{-i\omega_q \tau}-1 \big)\right] \label{eq:B}
\end{equation}
and the phase factor $\mathrm e^{i\phi}$. For the coupling of Eq.~(\ref{eq:gq}) the sum 
in Eq.~(\ref{eq:B}) may be performed analytically (cf. appendix).

The real and imaginary part of $B^-$ as functions of $\tau$ are shown in
Fig.~\ref{fig:B}. The real part exhibits an initial decay on the time scale
of about 1~ps and then saturates at a value slightly above 0.9. The imaginary
part is only nonzero during the initial decay of the real part. The
interference amplitude is the same function that describes the decay of the
optical polarization induced by the first laser
pulse.\cite{krummheuer2002theory} This decay is caused by the traveling wave
packet which after about 1~ps has left the QD resulting in an entanglement of
the QD with its environment and thus to a decoherence in the QD degrees of
freedom. Since the second pulse couples to the polarization of the QD
resulting from the first pulse, this decoherence reduces the coupling and
therefore also reduces the generation of phonons by this second pulse.

We want to mention that the saturation value of roughly 0.9 is also in good
agreement with experimental results obtained from time-integrated four-wave
mixing signals on ensembles of QDs, \cite{vagov2004non} where an initial decay
of the signals of about 20\% at a temperature of 4~K has been found.
According to the theory of four-wave mixing signals for the present type of
models \cite{vagov2002ele} this initial decay is given by
$1-|B^-(\infty)|^2$. This confirms the choice of the parameters used in our
present calculations as realistic ones.

\begin{figure}[ht]
\includegraphics[width=\columnwidth]{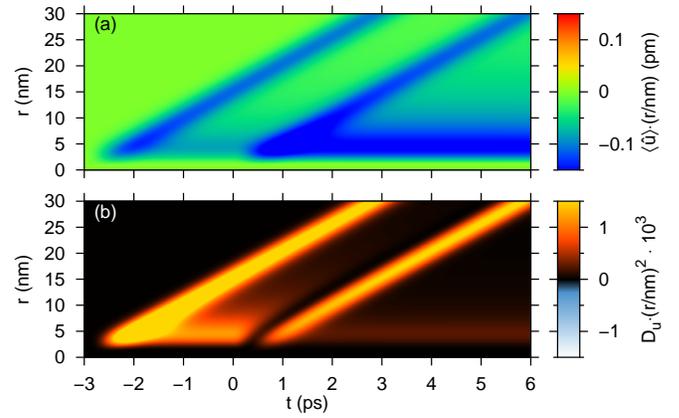}
\caption{\label{fig:2puls1} (Color online)
Same as Fig.~\ref{fig:1puls}, but after the excitation with two laser pulses
with pulse areas $\pi/2$ at $t=-3\,{\rm ps}$ and $t=0$ with a relative phase
of $\phi=0$.
}\end{figure}

Figure \ref{fig:2puls1} (a) shows the lattice displacement created after
excitation with a pair of $\pi/2$ pulses arriving at times $t=-3\,{\rm ps}$
and $t=0$ with a relative phase of $\phi=0$. We clearly see two emitted wave
packets, one starting at $t=-3\,{\rm ps}$ the other at $t=0$. At the time of
the second excitation the polaron amplitude is essentially doubled. For a
better interpretation of the results we note that for delay times longer than
$2\,{\rm ps}$ the interference amplitude $B^-$ is essentially real and its
value is larger than $0.9$. Therefore the amplitude of the displacement
created by the second pulse is not much smaller than the one created by the
first pulse. To simplify the discussion we set in the following $B^-\approx
1$, which corresponds to neglecting the dephasing of the polarization after
the first pulse. Note, however, that all the results shown in the Figures
have been calculated with the correct value of $B^-$.

Using $B^-\approx 1$,  Eq.~(\ref{eq:u2}) becomes
\begin{equation}
\langle \hat{u}(r,t)\rangle \simeq u_0 [ U(r,t+\tau) + \cos(\phi) U(r,t) ].
\label{eq:u2_approx}
\end{equation}
In the case of a phase difference $\phi=0$ as taken in Fig.~\ref{fig:2puls1}
the expectation value then evaluates to
\begin{equation}
\langle \hat{u}(r,t)\rangle\big|_{\phi=0} \simeq u_0 [2P(r) + W(r,t+\tau) + W(r,t)] ,
\end{equation}
where $P(r)$ and $W(r,t)$ are defined in Eqs.~(\ref{eq:u1}) and (\ref{eq:W}). 
This can be understood, when we look at the dynamics of the electronic
system. The second pulse excites the system from the equal superposition of
$|g\rangle$ and $|x\rangle$ to the exciton state $|x\rangle$. This doubling
of the exciton occupation causes a doubling of the polaron amplitude in the
QD. Because this process is similar to that induced by the first pulse, an
identical wave packet is emitted.

The fluctuations after a two pulse excitation with a large delay assuming $B^-\approx 1$ read
\begin{eqnarray}
D_u(r,t) &\simeq& \frac{u_0^2}{2} \big[ (W(r,t+\tau)  - W(r,t) )^2 \label{eq:Du2_approx}\\
&+& U(r,t)  (\sin^2(\phi)U(r,t)  + 2\sin(\phi)C(r,t;\tau)) \big] . \notag
\end{eqnarray}
In the fluctuations interferences between the single excitation processes
take place. They are summarized in the term $C(r,t;\tau)$. Its detailed form
is given in the appendix.

In Fig.~\ref{fig:2puls1}(b) we have plotted the corresponding fluctuations $D_u\cdot
r^2$. Positive values are seen at the positions of the two wave packets. The
horizontal line of the polaron, however, is almost missing after the second
pulse. These two features can directly be seen from the fluctuations which in this
case read
\begin{equation}
D_u(r,t)\big|_{\phi=0} \simeq \frac{u_0^2}{2} [W(r,t+\tau)^2 + W(r,t)^2] .
\end{equation}
Here we also assumed that $W(r,t+\tau)\cdot W(r,t)\approx 0$, because the
overlap of the two wave packets is negligible. The physical reason for the
absence of additional fluctuations in the QD beyond the vacuum fluctuations is the
fact that for a QD being completely in the exciton state the polaron
corresponds to a multimode shifted vacuum which has the same fluctuation
properties as the phonon ground state. So, in the present case the polaron
has fluctuations $D_u$ of almost zero and the positive fluctuations corresponding to the
two emitted wave packets remain. No squeezed states are thus created by this
laser pulse sequence.

\begin{figure}[ht]
\includegraphics[width=\columnwidth]{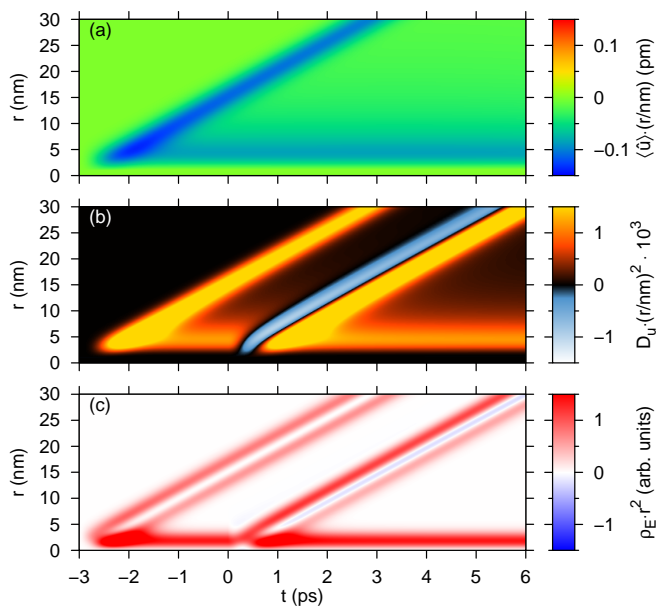}
\caption{\label{fig:2puls2} (Color online)
(a), (b) same as Fig.~\ref{fig:2puls1}, but for a relative phase $\phi=3\pi/2$. 
(c) energy density $\rho_E(r,t)\cdot r^2$.
}\end{figure}

From Eq.~(\ref{eq:Du2_approx}) one can see that squeezing can only occur if
$\sin(\phi)U(r,t)C(r,t;\tau)<0$. For $\phi=n\pi$ this term vanishes. Thus, if
squeezing occurs at all, we expect it to happen most prominently at phase
differences around odd multiples of $\pi/2$. In particular we will now study
the case $\phi=3\pi/2$ where $\sin(\phi)=-1$. Calculating the expectation
value of the displacement for this case, the outcome of
Eq.~(\ref{eq:u2_approx}) is
\begin{equation}
\langle \hat{u}\rangle\big|_{\phi=3\pi/2}=u_0 U(r,t+\tau),
\end{equation}
because $\cos(3\pi/2)=0$. In other words, the second excitation of the system
does not change the mean lattice displacement at all. Again looking at the
electronic system, the second pulse does not act on the occupation in the
electronic part of the system, we note that it just changes the relative phase in the
equal superposition of $|g\rangle$ and $|x\rangle$. This case is shown in
Fig.~\ref{fig:2puls2}(a), and indeed the expectation value looks exactly
like that of a single pulse excitation in Fig.~\ref{fig:1puls}(a).

It is interesting to note that according to the exact result for the
displacement given in Eq.~(\ref{eq:u2}), even in the case of arbitrarily
strong dephasing there is always a value of the relative phase given by
\begin{equation}
\tan(\phi) = \frac{\operatorname{Re}(B^-)}{\operatorname{Im}(B^-)}\label{eq:tanphi}
\end{equation}
where the second pulse does not modify the mean displacement.

In contrast, the fluctuations are strongly affected by the second
pulse, as can be seen in Fig.~\ref{fig:2puls2}(b). $D_u\cdot r^2$ clearly
shows fluctuations where the second wave packet would be expected. Thus our
analysis shows that there is an emission of a phonon wave packet also in this
case, however, with a vanishing mean displacement. We observe that in the
leading part of this wave packet clear negative parts build up in $D_u$. So a
squeezed phonon state is created under these excitation conditions.

Due to the non-zero fluctuations energy transport should occur even where the 
expectation value $\langle \hat{u}\rangle$ is zero. This energy can in principle be detected in 
bolometric measurements.\cite{hawker1999energy, bellingham2001acoustic} To 
quantify this, we calculate the energy density of the LA phonons given by
\cite{haken1976quantum}

\begin{eqnarray}
\rho_E({\bf r},t) &=& \frac{\hbar}{2V} \sum_{{\bf q},{\bf q}^\prime} 
\sqrt{\omega_q\omega_{q^\prime}} \left(1+\frac{\bf q}{q}\frac{{\bf q}^\prime}{q}\right) \\
&\times& \operatorname{Re} \left( \langle \hat{b}_{\bf q}^\dag\hat{b}_{{\bf q}^\prime}\rangle
 \mathrm e^{i({\bf q}-{\bf q}^\prime)\cdot{\bf r}} + \langle \hat{b}_{\bf q}\hat{b}_{{\bf q}^\prime}\rangle 
 \mathrm e^{i({\bf q}+{\bf q}^\prime)\cdot{\bf r}} \right) ,\notag
\end{eqnarray}
where the vacuum energy density has already been substracted. Again, after 
integrating over the Brillouin zone, the energy density only 
depends on the distance $r$ from the QD. We present $\rho_E(r,t) \cdot r^2$ in 
Fig.~\ref{fig:2puls2}(c). Corresponding to the emission of the first wave packet 
the energy density is non-zero around the diagonal line $(-3,0)\to (2,30)$. Also 
where the second wave packet is expected, namely at the diagonal line $(0,0)\to (6,30)$ 
the energy density is non-zero. Here it is mainly positive, and a small negative value appears, 
corresponding to an energy below the vacuum energy.

\begin{figure}[ht]
\includegraphics[width=\columnwidth]{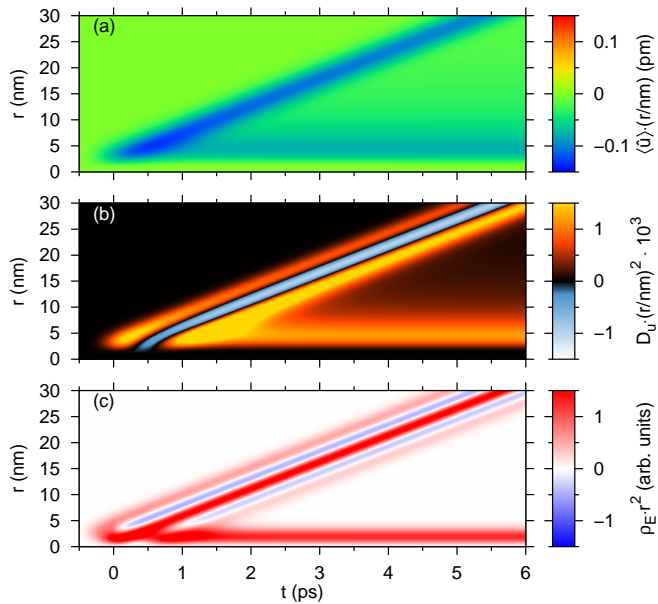}
\caption{\label{fig:2puls3} (Color online)
Same as in Fig.~\ref{fig:2puls2}, but with the first excitation
at $t=-0.5\,{\rm ps}$.
}\end{figure}

So far we have considered excitations, where the emitted wave packets are well
separated. When we reduce the delay to $\tau=0.5\,{\rm ps}$, the wave packets
created by the two laser pulses are overlapping in space. We take again a
relative phase of $\phi=3\pi/2$ which resulted in the appearance of squeezing
in the emitted wave packet. For this case $D_u$ is shown in
Fig.~\ref{fig:2puls3}(b). The mean displacement field is shown in Fig.~\ref{fig:2puls3}(a), 
it again does not exhibit much differences compared to the single pulse case.
This can be understood from the fact that also here $\operatorname{Im}(B^-)$
is much less than $\operatorname{Re}(B^-)$ and thus Eq.~(\ref{eq:tanphi}) is
still well satisfied for $\phi=3\pi/2$.

In the fluctuations $D_u$ we now find a broad wave packet with positive values in the
leading and the trailing part and pronounced negative values in between. The
interaction of the phonons created by the second pulse with the wave packet
resulting from the first pulse, which has not yet left the QD region,
enhances the squeezing almost by a factor of two compared to the case of 
separated wave packets.

The reduction of the delay time has a similar effect on the energy density, as is 
shown in Fig. \ref{fig:2puls3}(c). Here two significantly negative parts build up 
between the positive parts at the leading and trailing edges as well as at the center 
of the wave packets.

\begin{figure}[ht]
\includegraphics[width=0.7\columnwidth]{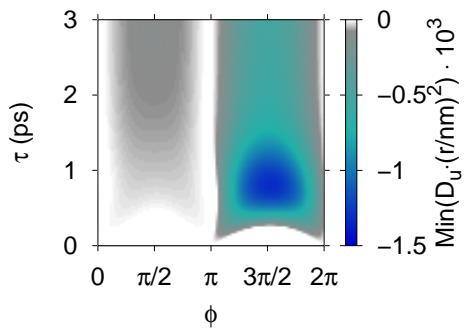}
\caption{\label{fig:phitau} (Color online)
Minimum values of $D_u\cdot r^2$ at $t=6\,{\rm ps}$ and $20\,{\rm nm}<r<50\,{\rm nm}$
as function of phase $\phi$ and delay $\tau$.
}\end{figure}

In order to obtain a complete picture of the squeezing behavior, when varying
either the delay time $\tau$ or the relative phase $\phi$ between the two
laser pulses, Fig.~\ref{fig:phitau} shows the minimum value of $D_u\cdot r^2$
in the emitted wave packets plotted as functions of $\tau$ and $\phi$. We
find that only relative phases in the range $\pi<\phi<2\pi$ lead to
remarkable squeezing. For $0<\phi<\pi$ the achievable squeezing values are
much smaller. When looking at the $\tau$-dependence of the plot, the largest
fluctuations $D_u$ are obtained for delay times between $0.5\,{\rm ps}$ and $1\,{\rm ps}$.
In this range the wave packets generated by the two laser pulses exhibit a
spatial overlap which obviously favors the build up of squeezing. For delays
$\tau\approx1\,{\rm ps}$ the two wave packets have a spatial distance of $r=c
\tau\approx 5\,{\rm nm}$, which is approximately the size of a single wave
packet. For larger delays the spatial overlap is small and the squeezing
values are reduced.

\section{Conclusions}\label{s:conclus}

In conclusion, we have analyzed the fluctuation properties of LA phonons
after ultrafast excitation of a QD in resonance with the lowest exciton transition. 
Like in the case of LO phonons,\cite{sauer2010, reiter2011wig} also for LA phonons 
we have found that a single pulse never gives rise to squeezing, but a sequence of two such
excitations can create squeezed phonons. In contrast to the optical phonons,
which due to their vanishing group velocity remain confined in the QD, the
excitation of LA phonons leads to the formation of wave packets that leave
the QD. We have shown that the phononic wave packet emitted after the
excitation by the second pulse can be squeezed. In contrast to the squeezing found for 
LA phonons generated from the decay of LO phonons\cite{daniels2011quantum} here 
also the phonon wave packets that leave the dot exhibit squeezing. The appearance 
of squeezing depends crucially on the relative phase $\phi$ between the two pulses.
Sizeable squeezing is only found in the range $\pi<\phi<2 \pi$. For short delay
times, when there is an overlap between the wave packets generated by the two
pulses, the interaction between these wave packets has a significant impact
on the strength of the fluctuations.

Let us finally comment on the achieved values of the squeezing, which at
first seems to be rather small. However, it should be noted that vacuum
fluctuations at a given point result from all phonon modes in the first
Brillouin zone. Excitation induced fluctuations, on the other hand, occur
only for those phonon modes which couple to the QD exciton. According to the
coupling matrix element of Eq.~(\ref{eq:gq}) these are only phonons with
vectors up to about the inverse size of the QD. If we would consider in the
calculation of $D_u$ only those phonon modes which couple to the QD
exciton a rather pronounced squeezing would emerge. Such a reduction of the
effective range of phonon modes occurs also in an experiment if the
measurement of the fluctuations is performed with a finite spatial
resolution, which again eliminates the fluctuations resulting from phonons
with wave vectors larger than the inverse spatial measurement
resolution.\cite{papenkort2012optical}

\appendix

\section{}
For completeness here we show the full analytical forms of some quantities
introduced in the main text. First the fluctuations for the two pulse excitation from
Eq.~(\ref{eq:Du2_approx}) without approximation read
\begin{eqnarray}
D_u(r,t) &=& \frac{u_0^2}{2} \big[ (W(r,t+\tau)  - W(r,t) )^2 \label{eq:Du2}\\
&+& U(r,t)  ((1-R^2)U(r,t)  + 2I C(r,t;\tau)) \big] \notag
\end{eqnarray}
with $R=\operatorname{Re}(B^-\mathrm e^{i\phi})$ and
$I=\operatorname{Im}(B^-\mathrm e^{i\phi})$. Here and in Eq.~(\ref{eq:B}) the
interference amplitude appears; its full form is:
\begin{eqnarray}
B^-&=&\exp\left[\frac{V}{2\pi^2}\int_0^\infty q^2
\left|\frac{g_q}{\omega_q}\right|^2 \big( \mathrm e^{-{\rm i}\omega_q \tau}-1
\big)\,{\rm d}q \right]\\
&=&\exp\left[-\frac{\tau}{4\pi^{3/2}\varrho\hbar c^2}\times\right.\\
&&\left\{\frac{{D^e}^2}{2a_e^3}\left[i+\operatorname{erfi}\left(\frac{c\tau}{a_e}\right)\right] 
+ \frac{{D^h}^2}{2a_h^3}\left[i+\operatorname{erfi}\left(\frac{c\tau}{a_h}\right)\right] \right.\notag\\
&&\left.\left.\qquad -2 \frac{D^eD^h}{(a_e^2+a_h^2)^{3/2}}
\left[i+\operatorname{erfi}\left(\frac{c\tau}{\sqrt{a_e^2+a_h^2}}\right)\right]\right\}\right] \notag
\end{eqnarray}
with the complex error function $\operatorname{erfi}(x)=i\cdot\operatorname{erf}(ix)$. 
The interference term itself from Eqs.~(\ref{eq:Du2}) and
(\ref{eq:Du2_approx}) is: 
\begin{eqnarray*}
C(r,t;\tau) &=& \tilde{C}(r-ct) + \tilde{C}(r+ct)\\
&-&[\tilde{C}(r-c(t+\tau)) + \tilde{C}(r+c(t+\tau))]
\end{eqnarray*}
with

\begin{eqnarray*}
\tilde{C}(r-ct) &=& \frac{\sqrt{\pi}}{r} \left[\frac{D^e}{a_e}
\operatorname{erfi}\left(\frac{r-ct}{a_e}\right) \mathrm e^{-\frac{(r-ct)^2}{a_e^2}} \right.\\
&&\quad \left. - \frac{D^h}{a_h} \operatorname{erfi}\left(\frac{r-ct}{a_h}\right)
\mathrm e^{-\frac{(r-ct)^2}{a_h^2}} \right] \notag \\
&+&\frac{1}{2r^2} \left[ \frac{D^e}{a_e^2} H\left(- \frac{(r-ct)^2}{a_e^2}\right) \right. \\
&&\quad \left. - \frac{D^h}{a_h^2} H\left(- \frac{(r-ct)^2}{a_h^2} \right) \right] (r-ct)^2 .\notag \\
\end{eqnarray*}

and
\begin{equation*}
H(x) = {}_2F_2 \left(1,1\,;\,\frac32,2\,;\,x\right) =
\sum_{n=0}^\infty \frac{2x^n}{(2n+1)!!\cdot(n+1)} 
\end{equation*}
is a generalized hypergeometric function.

The complete analytical form of the energy density after integration over $q$ and $q^\prime$ is

\begin{eqnarray}
\rho_E(r,t) &=& \frac{1}{64\pi^4\varrho c^2} [ \mathcal E(r,t)^2 + \mathcal E(r,t+\tau)^2 \label{eq:rhoE}\\
&+& \mathcal E(r,t)((R-1)\mathcal E(r,t+\tau) - I \tilde{\mathcal E}(r,t;\tau)) ] , \notag\\
\mathcal E(r,t) &=& E_1(r-ct) + E_1(r+ct) -2 E_1(r) \notag\\
&+& E_2(r-ct) - E_2(r+ct) ,\notag\\
\tilde{\mathcal E}(r,t) &=& \tilde{\mathcal E}_1(r,t;\tau) + \tilde{\mathcal E}_2(r,t;\tau),\notag\\
\tilde{\mathcal E}_1(r,t;\tau) &=& \tilde{E}_1(r-ct) + \tilde{E}_1(r+ct) \notag\\
&-& \tilde{E}_1(r-c(t+\tau)) - \tilde{E}_1(r+c(t+\tau)) ,\notag\\
\tilde{\mathcal E}_2(r,t;\tau) &=& \tilde{E}_2(r-ct) + \tilde{E}_2(r+ct) \notag\\
&-& \tilde{E}_2(r-c(t+\tau)) - \tilde{E}_2(r+c(t+\tau))\notag
\end{eqnarray}
\begin{eqnarray*}
E_1(r-ct) &=& \frac{2\sqrt{\pi}}{r}  \left(\frac{D^e}{a_e^3}\mathrm e^{\frac{(r-ct)^2}{a_e^2}} 
- \frac{D^h}{a_h^3}\mathrm e^{\frac{(r-ct)^2}{a_h^2}} \right) (r-ct)\\
E_2(r-ct) &=& E_1(r-ct)\\
&+& \frac{\sqrt{\pi}}{r^2} \left( \frac{D^e}{a_e}\mathrm e^{-\frac{(r-ct)^2}{a_e^2}} 
- \frac{D^h}{a_h}\mathrm e^{-\frac{(r-ct)^2}{a_h^2}}\right)\\
\tilde{E}_1(r-ct) &=& \frac{2\sqrt{\pi}}{r}\left( \frac{D^e}{a_e^3}\operatorname{erfi}
\left(\frac{r-ct}{a_e}\right)\mathrm e^{-\frac{(r-ct)^2}{a_e^2}} \right.\\
&&\quad \left. - \frac{D^h}{a_h^3}\operatorname{erfi}\left(\frac{r-ct}{a_h}\right)
\mathrm e^{-\frac{(r-ct)^2}{a_h^2}} \right) (r-ct)\\
\tilde{E}_2(r-ct) &=& \tilde{E}_1(r-ct) \\
&+& \frac{\sqrt{\pi}}{r^2} \left( \frac{D^e}{a_e}\operatorname{erfi}\left(\frac{r-ct}{a_e}\right)
\mathrm e^{-\frac{(r-ct)^2}{a_e^2}} \right. \\
&&\qquad \left. - \frac{D^h}{a_h}\operatorname{erfi}\left(\frac{r-ct}{a_h}\right)
\mathrm e^{-\frac{(r-ct)^2}{a_h^2}} \right)
\end{eqnarray*}

\end{document}